**Mirror symmetry breaking in micelles of N-stearoyl serine enantiomers**

Yosef Scolnik

IYAR (Israel Institute for Advanced Research); Weizmann Institute of Science

yosef.scolnik@weizmann.ac.il

*Dedicated to the memory of Proff. Meir Shinitzky - his path I follow, on his shoulders I stand.

Non-identity of enantiomers, in contrary to the existing dogma, is proved by this work

**Abstract:** Circular dichroism spectra were recorded for micellar aggregates of N-stearoyl (L or D) serine in $H_2O$ or $D_2O$. Micelle formation kinetics differed markedly in $H_2O$, but in $D_2O$ the enantiomers showed similar spectral characteristics. The results confirm previous observations (*1*) that described differences in the thermodynamic properties of enantiomers, in contrary to the reigning dogma. The comparison of spectral properties indicates that this phenomenon depends on the interactions with $H_2O$ and is not due to trivial contamination.

Main

The origin of biological homochirality incorporating L-amino acids and D-sugars remains an unanswered question that has been discussed in numerous monographs and reviews (*2-12*). One approach is that whatever the cause, the presumed identity of the physical properties of enantiomers make spontaneous symmetry breaking a matter of chance, so that life could equally well have incorporated D-amino acids and L-sugars (*3*).

An alternative view posits that symmetry breaking is deterministic. For instance, it is now established that left–right symmetry is broken at the level of interactions between subatomic particles and atoms, and it is possible that this fundamental dissymmetry can be transferred to the molecular scale (*2*). The spectral characteristics and a crucial comparison with $D_2O$ strongly support the existence of significant differences between enantiomers in their interactions with $H_2O$, but not with $D_2O$.

**Methods**

*Preparation of stearoyl serine solutions*

Stearic acid, N-hydroxysuccinimide, ethyl acetate, dicyclohexylcarbodiimide, ethanol, tetrahydrofuran, L and D serine, sodium bicarbonate, hydrochloric acid and acetone were purchased from Sigma Aldrich. The chiral surfactants, D- and L-stearoyl serine, (D-NSS and L-NSS) were prepared by coupling stearic acid N-hydroxy succinimide ester with D and L serine, as described by Lapidot et al (*11*). The structures and compositions of the D-NSS and L-NSS were confirmed by $^1$H-NMR, mass spectroscopy, infrared spectra and optical rotation. Mass spectroscopy showed that the purity of the D and L stearoyl serine compounds exceeded 98.5%.

Initial solutions of L-NSS and D-NSS were prepared in 1:1 ethanol:water at a concentration of 0.1 M. Their concentrations, calibrated by absorption spectra, were found to be identical within the limits of instrument resolution. An aliquot of the respective initial solution was mixed 1:1000 (18 µL into 1800 µL) in $H_2O$ and $D_2O$ (0.01 M KOH solutions) to form a working solution $10^{-4}$ M of L-NSS and D-NSS.

*Circular dichroism measurements*

CD measurements using a 1 mm quartz cuvette were performed with a Chirascan spectrophotometer (Applied Photophysics) from 280 to 196 nm in 1 nm steps, 3 seconds per point, and 4 minutes between successive runs. Measurements began two minutes after the working solutions were prepared, and the baseline of the solvent was subtracted.

*Controls for impurities*

To determine whether impurities present in the water were somehow affecting the measurements, the double distilled water (DDW) was evaporated until 1/20 of the initial volume remained, so that any impurities would be concentrated 20-fold. The residual water was then used to prepare and monitor micelle formation. The results were identical to those with DDW, weighing against the possibility that unknown contaminants in the water were causing the observed differences.

**Results**

Figure 1 shows the CD spectra of the two enantiomers in 1:1 methanol/water KOH 0.01M, which are conditions that suppress micelle formation. The spectra are identical, confirming that the differences observed in CD spectra when micelles are able to form are not a consequence of impurities or variations in the way the solutions are prepared.

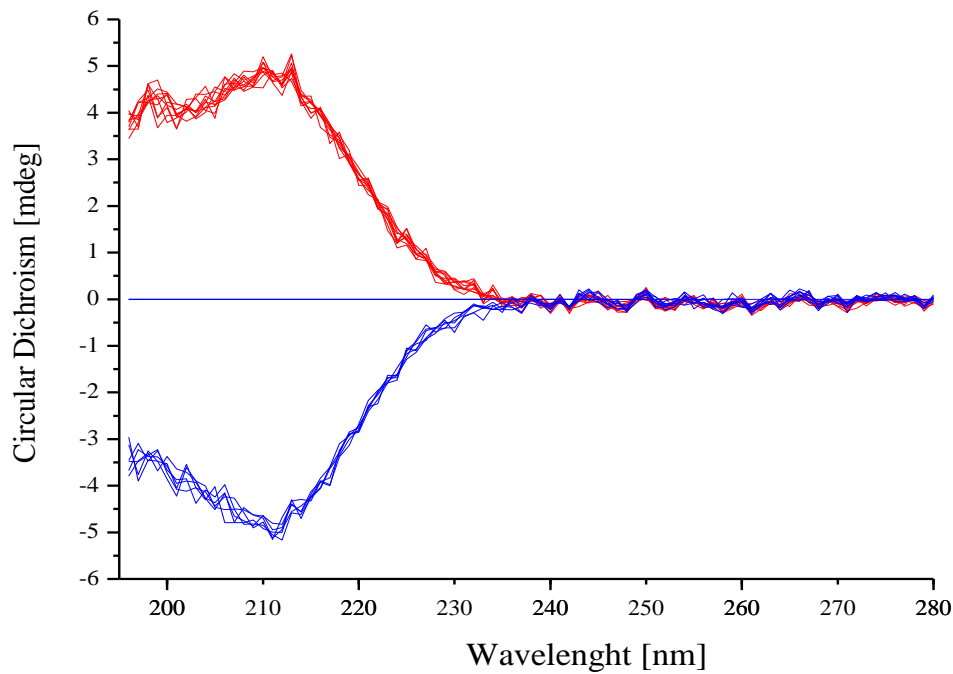

**Figure 1.** Circular dichroism of D-NSS (D-stearoyl serine, upper curves) and L-NSS (L-stearoyl serine, lower curves) in 0.01M KOH in $H_2O$, mixed 1:1 with methanol. These conditions inhibit micelle formation and the spectra are identical.

When conditions conducive to micelle formation are employed, there is a clear difference in the rates of micelle formation between L-NSS and D-NSS in $H_2O$, as depicted in Figure 2. The spectra of L-NSS micelles change only slightly over time, as shown by the repetitions, to about 32 millidegrees. In contrast, the spectra of D-NSS micelles in $H_2O$ grow markedly over time and at a crucial size change in shape. The magnitude of D-NSS is also higher, at approximately -62 millidegrees.

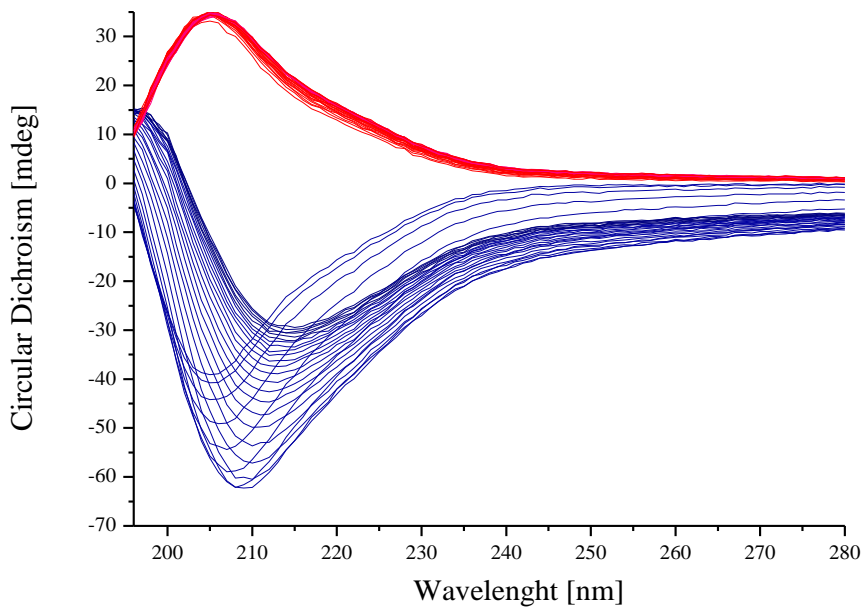

**Figure 2.** Circular dichroism of L-NSS (L- stearoyl serine, upper curves) and D-NSS (D-stearoyl serine, lower curves) in 0.01M KOH in $H_2O$. These conditions promote micelle formation. Each spectrum is separated from the next by four minutes.

In $D_2O$ the spectra of D-NSS resembles its spectra in $H_2O$ spectra as illustrated in Figure 3. However, the spectra of L-NSS are markedly different from those in $H_2O$ and now more closely resemble the D-NSS spectra.

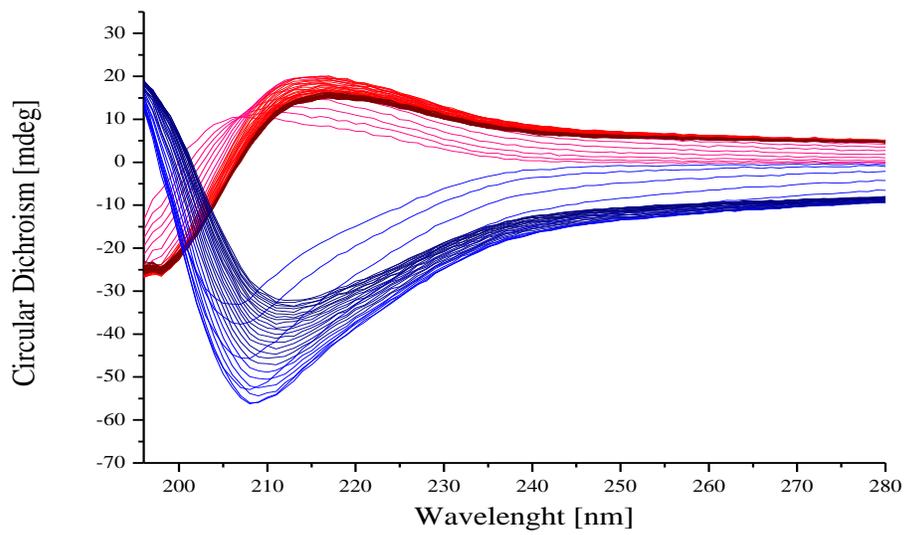

**Figure 3**. Circular dichroism of L-NSS (L-stearoyl serine, upper curves) and D-NSS (D-stearoyl serine, lower curves) in 0.01M KOH in $D_2O$. These conditions promote micelle formation. Each spectrum is separated from the next by four minutes.

In order to illustrate the differences in kinetic behavior of L-NSS and D-NSS, time points were plotted for the spectral shifts of the two enantiomers in $H_2O$ and $D_2O$ as depicted in Figure 4. The L-NSS reaches its equilibrium value within two minutes in $H_2O$, but more slowly in $D_2O$, whereas D-NSS' kinetic behavior is similar in $H_2O$ and $D_2O$ and requires significantly more time to reach equilibrium.

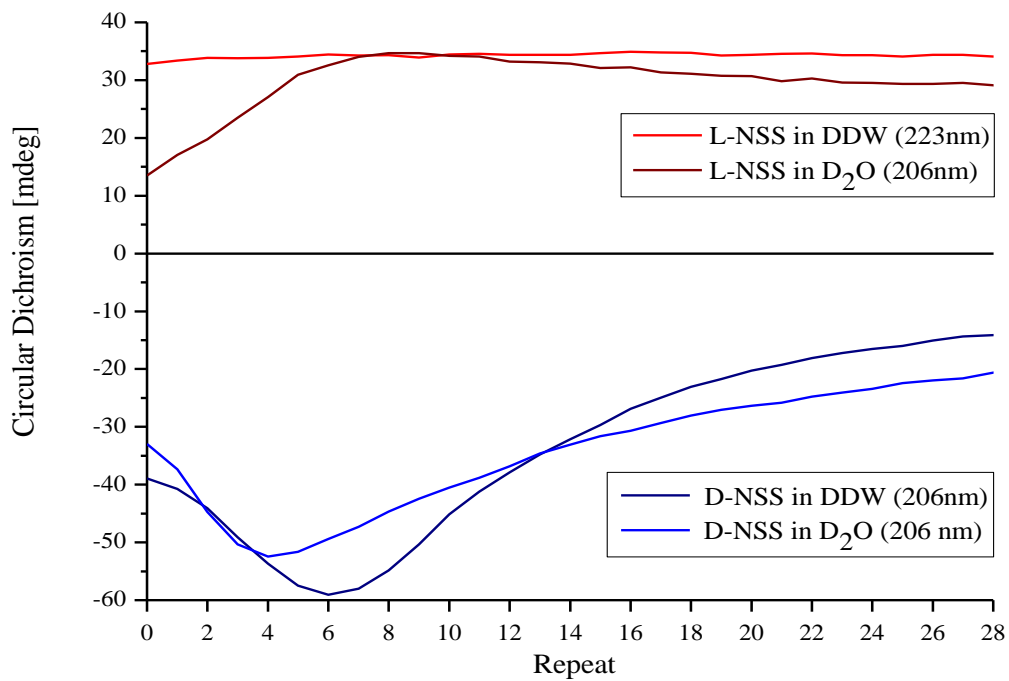

**Figure 4.** Circular dichroism of L-NSS (L-stearoyl serine) in $H_2O$ (223 nm), L-stearoyl serine) in $D_2O$ (206 nm), D-NSS (D-stearoyl serine) in $H_2O$ (206 nm) and D-NSS (D- stearoyl serine) in $D_2O$ (206 nm), versus time (represented by repeating the measurement every four minutes).

The identity in structure as seen in the CD spectra between D-NSS micelles in DDW and in $D_2O$ can be visualized using electron microscopy, as clearly shown in Figure 5. Furthermore, the shift in structure of L-NSS micelles in $D_2O$, compared to the structure in DDW, to identity with the D-NSS structure, is illustrated. D-NSS micelles begin forming a structure resembling the L-NSS micelles in DDW and a shift in structure occurs during formation in agreeance with the CD spectra. The difference in structure between L-NSS micelles in DDW and D-NSS micelles in DDW is distinct.

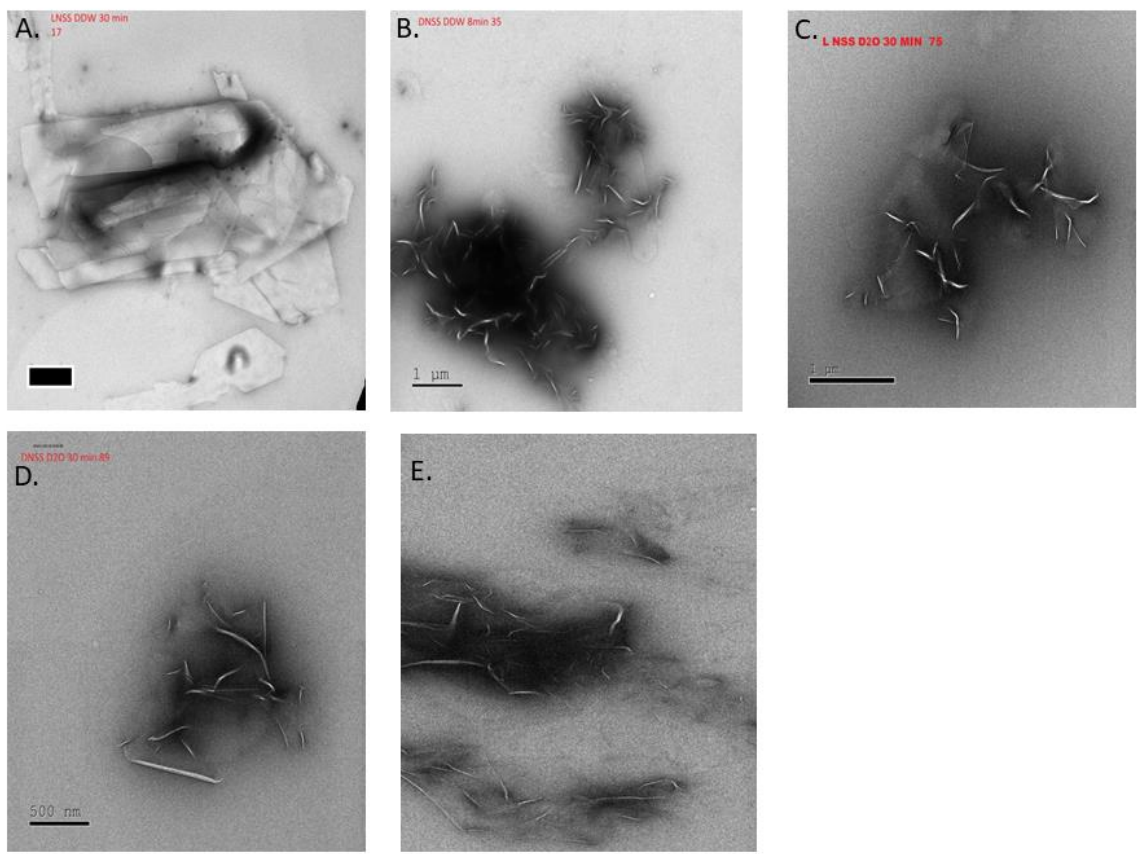

**Figure 5.** Electron micrographs of L-NSS and D-NSS micelles in DDW. The micelles were prepared by the same technique described in Methods. Briefly, an aliquot from the respective solutions were sampled 8 and 30 minutes after beginning of the micellular formation. Specimens were prepared for electron microscopy using the standard staining method. A. L-NSS micelles after 30 minutes; B. D-NSS micelles formation after 8 minutes (at the right upper side of the picture remnants of structures akin to the L-NSS, can be seen); C. L-NSS micelles in $D_2O$ after 30 minutes. The shift from the structure in DDW to a structure identical to D-NSS micelles structure (in DDW and $D_2O$) is shown; D. D-NSS micelles in $D_2O$ after 30 minutes; E. D-NSS micelles in $D_2O$ after 30 minutes.

**Discussion**

Studies from the Shinitzky group have suggested that hydration shells around chiral amino acids in water could enhance chiral discrimination (*1, 13*). In the case of polyaminoacids and peptides dissolved in water, differences between poly-L and poly-D peptides were observed in helix-coil transition (*1*) as well as in thermal stability and supramolecular structures (*14*). When tested in $D_2O$ under identical conditions these differences were markedly reduced, indicating a specific discrimination between chiral configurations by $H_2O$ molecules. It has been proposed (*1, 13*) that the spin isomers of water, i.e. ortho-$H_2O$ and para-$H_2O$, present in a 3:1 ratio could account, at least partially, for this discriminatory power. In $D_2O$ the spin isomers are much less pronounced, which drastically attenuate the putative chiral preference.

Several publications from other investigators support this notion. In 2008, Kodona et al reported that the α-helical motif in poly-D-glutamic acid is stable thermally, compared to that of poly-L-glutamic acid. This difference was ascribed to an amplified interaction of the L-species with the more abundant ortho-water (*14*). Guo et al investigated the dilution enthalpies of enantiomers of six beta-amino alcohols in dimethylsulfoxide (DMSO) + $H_2O$ mixtures using an isothermal titration calorimeter, and calculated the corresponding homochiral enthalpic pairwise interaction coefficients (hXX) of the six amino alcohols. It was found that throughout the composition range of mixed solvents, values of hXX for the S-enantiomer are virtually all greater than those of R-enantiomer for each amino alcohol (*15*). Fujiki et al investigated several ambidextrous artificial helical Si–Si bonded polysilanes in solution as a function of temperature. Their preliminary results indicated subtle differences in chiroptical and achiral $^{29}$Si-NMR and viscometric data for a pair of helical polysilanes (*16*).

*Can parity-violating energy difference (PVED) account for these variances?*

Theoretical studies (*2, 4*) have established that an image and its mirror molecule are not true enantiomers due to a parity-violating energy shift (PVES) of the electronic binding energy with a positive value, $^+$EPV, for the image molecule and a PVES with a negative value, $^-$EPV, for the mirror image molecule (or vice versa); PVES makes the pair become a diastereomer. The parity-violating energy difference (PVED) between image and mirror image molecules is given as PVED = EPV – (–EPV) = 2•EPV.

Other studies have proposed that PVES might be detectable as slight differences in electronic transition energies (*2, 17*). However, because of the extremely small PVED of ~$10^{-19}$ eV, it is generally believed that parity violation cannot induce observable macroscopic thermodynamic differences between enantiomers. To overcome the vanishingly small PVED, several amplification models have been proposed to explain the emergence of biological homochirality (*2*).

*Stearoyl serine enantiomer properties are different in $H_2O$ and $D_2O$*

Micelle formation of L and D-stearoyl serine and the differences between the enantiomers were discovered by Shinitzky et al (*18*). N-stearoyl serine micelles presumably originate from a repetitive arrangement of the amide planes on the micellar surface, forming a "quasi- peptide" bonded by hydrogen bonds, replacing peptide bonds.

The formation of the micelles by hydrogen bonds instead of strong peptide bonds makes it a very sensitive system that amplifies the influence of solvents.

The spectral kinetics of micelle formation reported here provide further support for the notion that enantiomers have non-identical thermodynamic macroscopic properties, perhaps due to parity violation amplification by the differential interaction of the enantiomers with $H_2O$. The observation that the differences are considerably reduced in $D_2O$ supports this hypothesis. This putative chiral preference in $H_2O$ may encapsulate the basic molecular mechanism of parity violation amplification which is capable of transferring PVED to the molecular level. Differential hydration of enantiomers may also have played a crucial role in the deterministic selection of L-amino acids by early forms of life.

**References**

1       Scolnik, Y. *et al.* Subtle differences in structural transitions between poly-L- and poly-D-amino acids of equal length in water. *Phys Chem Chem Phys* **8**, 333-339, doi:10.1039/b513974k (2006).
2       MacDermott, A. Perspective and concepts: Biomolecular Significance of Homochirality: The Origin of the Homochiral Signature of Life  *Comprehensive Chirality* **8**, 11-37 (2012).
3       Lahav, M. e. a. Stochastic "Mirror Symmetry Breaking" via Self-Assembly, Reactivity and Amplification of Chirality: Relevance to Abiotic Conditions. *Top Curr Chem* **259**, 123–165 (2005).
4       Guijarro, A. & Yus, M. in *The Molecules of Life* (RSC, 2009).
5       Ball, R. & Brindley, J. The Life Story of Hydrogen Peroxide III:Chirality and Physical Efforts at the Dawn of Life *Origin of Life Evolution Biosphere* **46**, 81-93 (2016).
6       Konstantinov, K. K. & Konstantinova, A. Chiral Symmetry Breaking in Peptide Systems During Formation of Life on Earth. *Origin of Life Evolution Biosphere* **48**, 93-122 (2018 ).
7       Tverdislov, V. A. [Chirality as a primary switch of hierarchical levels in molecular biological systems]. *Biofizika* **58**, 159-164 (2013).
8       Berger, R., Quack, M. & Tschumper, G. S. Electroweak quantum chemistry for possible precursor molecules in the evolution of biomolecular homochirality. *Helvetica Chimica Acta* **83**, 1919-1949 (2000).

9       Plasson, R., Kondepudi, D. K., Bersini, H., Commeyras, A. & Asakura, K. Emergence of homochirality in far-from-equilibrium systems: mechanisms and role in prebiotic chemistry. *Chirality* **19**, 589-600, doi:10.1002/chir.20440 (2007).

10      Ribo, J. M., Hochberg, D., Crusats, J., El-Hachemi, Z. & Moyano, A. Spontaneous mirror symmetry breaking and origin of biological homochirality. *J R Soc Interface* **14**, doi:10.1098/rsif.2017.0699 (2017).

11      Lapidot, Y., Rappoport, S. & Wolman, Y. Use of esters of N-hydroxysuccinimide in the synthesis of N-acylamino acids. *J Lipid Res* **8**, 142-145 (1967).

12      Kondepudi, D. K. & Nelson, G. W. Weak neutral currents and the origin of biomolecular chirality. *Nature* **314** 438–441 (1985).

13      Deamer, D. W., Dick, R., Thiemann, W. & Shinitzky, M. Intrinsic asymmetries of amino acid enantiomers and their peptides: a possible role in the origin of biochirality. *Chirality* **19**, 751-763, doi:10.1002/chir.20434 (2007).

14      Kodona, E. K., Alexopoulos, C., Panou-Pomonis, E. & Pomonis, P. J. Chirality and helix stability of polyglutamic acid enantiomers. *J Colloid Interface Sci* **319**, 72-80, doi:10.1016/j.jcis.2007.10.063 (2008).

15      Liang, H. *et al.* Pairwise interaction enthalpies of enantiomers of beta-amino alcohols in DMSO + H2O mixtures at 298.15 K. *Chirality* **24**, 374-385, doi:10.1002/chir.22021 (2012).

16      Fujiki, M. Mirror symmetry breaking of silicon polymers--from weak bosons to artificial helix. *Chem Rec* **9**, 271-298, doi:10.1002/tcr.200900018 (2009).

17      Zel'dovich, B. Y., Saakyan, D. & Sobel'man, I. Energy difference between right- and left-hand molecules, due to parity nonconservation in weak interactions of electrons with nuclei. *Sov. Phys. JETP. Lett.* **25**, 94–97 (1977).

18      Shinitzky, M. & Haimovitz, R. Chiral surfaces in micelles of enantiomeric N-palmitoyl- and N-stearoylserine. *J. Am. Chem. Soc.* **115**, 12545-12549 (1993).

**Acknowledgments**

The author thanks to the Kimmelman Center for Biomolecular Structure and Assembly, for financial support to this work.

I would like to thank Yechiel Goldman for editorial assistance and Tanya Lasitza-Male for her helps and support.